\documentclass[12pt,preprint]{aastex}

\shortauthors{Cowan \& Ivezi\' c}
\shorttitle{The Environment of Galaxies}
\begin{document}

\title{The Environment of Galaxies at Low Redshift}

\author{Nicolas B. Cowan\altaffilmark{1},
	 \v Zeljko Ivezi\' c\altaffilmark{1},
	 }

\altaffiltext{1}{Astronomy Department, University of Washington,
   Box 351580, Seattle, WA  98195\\
email: cowan@astro.washington.edu, ivezic@astro.washington.edu}

\begin{abstract}
We compare environmental effects in two analogous samples of galaxies, one from the Sloan Digital Sky Survey (SDSS) and the other from a semi-analytic model (SAM) based on the Millennium Simulation (MS), to test to what extent current SAMs of galaxy formation are reproducing environmental effects. We estimate the large-scale environment of each galaxy using a Bayesian density estimator based on distances to all ten nearest neighbors and compare broad-band photometric properties of the two samples as a function of environment. The feedbacks implemented in the semi-analytic model produce a qualitatively correct galaxy population with similar environmental dependence as that seen in SDSS galaxies. In detail, however, the colors of MS galaxies exhibit an exaggerated dependence on environment: the field contains too many blue galaxies while clusters contain too many red galaxies, compared to the SDSS sample. We also find that the MS contains a population of highly clustered, relatively faint red galaxies with velocity dispersions comparable to their Hubble flow. Such high-density galaxies, if they exist, would be overlooked in any low-redshift survey since their membership to a cluster cannot be determined due to the ``Fingers of God'' effect.
\end{abstract}

\keywords{\
galaxies: fundamental parameters ---
galaxies: luminosity function, mass function ---
galaxies: statistics ---
}

\section{Introduction}
Since the discovery of the morphology-environment relation \citep{Dressler_1980}, it has been known that galaxy properties are correlated with their large-scale environment: the average morphology, color and luminosity of galaxies differ depending on how crowded their neighborhood is. On the face of it, it is not clear why or how the environment of a galaxy on Mpc scales should be related to the kpc-scale processes (star formation, supernova and AGN feedback) that determine the bulk properties of a galaxy. To further confuse matters, the strong correlation between morphology, color and luminosity \citep[][and references therein]{Strateva_2001} makes it unclear which property is ultimately driven by environment let alone which physical processes are responsible. It is not even clear to what extent the environment of a galaxy effects it through nature (different formation conditions) rather than nurture (galaxy-galaxy interactions). A critical step towards answering such questions is to compare observed trends to those present in a simulated galaxy ensemble in which one knows all the processes at work. 

The Sloan Digital Sky Survey \citep[SDSS,][]{Adelman_2007} is a powerful tool for addressing questions of environmental effects. Its spectroscopic sample of galaxies is the largest such sample ever, ensuring that even relatively rare galaxy populations are well represented, and the survey's large contiguous footprint makes it easy to determine the large-scale environment for most of these galaxies. Previous researchers who have used the SDSS galaxy catalog to study environmental dependences have found three broad trends: the peaks of the bimodal color distribution of galaxies do not shift for different environments; blue and red galaxies are most common in low- and high-density environments, respectively; the luminosity of red galaxies increases with local density \citep{Hogg_2003,Kauffmann_2004,Balogh_2004,Tanaka_2004,Blanton_2005,Zehavi_2005,Park_2007, Ball_2007}.

The Millennium Simulation \citep[MS,][]{Springel_2005} is the largest ever cosmological simulation comprising some $10^{10}$ dark matter (DM) particles with a spatial resolution of $5h_{100}^{-1}$ kpc. At $z=0$, the simulation fills a cube $500h_{100}^{-1}$ Mpc per side. The MS does not explicitly model the gas, dust and stars which make up observable galaxies, but it produces a DM halo merger tree which serves as the backbone for a number of semi-analytic models (SAMs). Unlike N-Body/SPH simulations, SAMs do not simulate the gravitational and hydrodynamic forces involved in the formation and evolution of galaxies, but they do provide a computationally inexpensive way to explore the parameter space of sub-grid processes. The trade-off is that the parameters of a SAM must be tuned using observations (e.g. matching to the observed luminosity function), making truly independent comparisons between the model and reality more challenging. Numerous groups have developed SAMs which hierarchically form some $10^{7}$ galaxies from the MS merger tree  \citep[eg:][and references therein]{Croton_2006, Bower_2006, DeLucia_2007}. Their models differ ---for example in their treatment of AGN feedback--- but all reproduce some of the empirical features of galaxy populations. The MS galaxies have a very realistic distribution of luminosities, thanks to judicious use of ``radio'' feedback. They also exhibit a bimodal color distribution as discovered in SDSS \citep{Strateva_2001}. Finally, the power spectrum of the density fluctuations is in good agreement with the empirical data from 2dF and SDSS \citep{Springel_2005}. Previous investigators have found that the brightest MS galaxies are red, dead ellipticals populating rich galaxy clusters \citep{DeLucia_2006}, while the modeled galaxies in the very lowest-density environments have similar colors and star formation rates as analogous SDSS galaxies \citep{Patiri_2006}. 

In this work we compare the observed galaxy populations with those produced with SAMs. Our work differs from those listed above in the following ways: we use Bayesian number density as a proxy for local environment, rather than the commonly used surface density or two-point correlation function; we use the $u-r$ color, which has more leverage than the $g-r$ color; we use SDSS Data Release 5, rather than any of the previous (smaller) releases; and last but not least, we compare observed and modeled galaxies for the full range of galaxy environments and colors.

\section{Selection Criteria}
We use a sample of 674,749 galaxies from the SDSS Data Release 5 (DR5) main galaxy sample, an optical imaging and spectroscopic survey of galaxies over 1/4 of the sky (with limiting magnitude $r < 17.7$ after foreground extinction removal). We create a complete volume and luminosity-limited sample with $0.01 < z < 0.077$ (or distances of 43--345~Mpc from the Milky Way) and $M_{r} < -20$, leaving 90,689 galaxies. The characteristic galaxy luminosity in the SDSS r-band is $M_{*} = -20.60$ \citep{Blanton_2003b}, which falls well within our magnitude limit. We use model magnitudes corrected for foreground extinction but do not apply K corrections (we instead apply K corrections to the model galaxies); absolute magnitudes are computed using $h_{0}=0.732$ \citep{Spergel_2007}. As our sample of SDSS galaxies only extends to lookback times less than 1~Gyrs, it is representative of local galaxies.

We use the modeled galaxies generated by \cite{DeLucia_2007} and available online through the Millennium Simulation database\footnote{www.g-vo.org/MyMillennium}. The quantities we use are the Cartesian positions and velocities of the galaxies, as well as their absolute SDSS $u$ and $r$ magnitudes, which include dust extinction from both a diffuse ISM and attenuation of stars in young clusters in the emitting galaxy \citep{DeLucia_2006}. We make a $M_{r} < -20$ cut on the $z=0$ snapshot, resulting in a complete sample of 1,805,780 galaxies in the simulation volume. We create a mock observational catalog for an observer at the origin by computing the right ascension, declination and distance modulus of each galaxy. Since the model galaxies have rest-frame colors, we apply K corrections using the model spectra of \cite{Bruzual_2003}, although this changes the colors by less than $0.2$ magnitudes. We then make cuts on radial distance, keeping only those galaxies which fall within the 43--345~Mpc range of our volume-limited SDSS sample. This mock ``survey'' covers one eighth of the sky and contains 110,437 galaxies, $\sim 20$\% more than our SDSS sample.   

\section{Bayesian Density Estimator}
The most common proxy for environment is the number density of galaxies, or the surface density of galaxies within redshift slices. For example, \cite{Blanton_2005} use a deprojected angular correlation function, while \cite{Scoville_2007} use an adaptively smoothed surface-density. Other groups have used three-dimensional density estimators, such as over-density on a $8h^{-1}$ Mpc scale \citep{Hogg_2003} or within a smoothing kernel \citep{Park_2007}. \cite{Mateus_2007} use a hybrid of three-dimensional and two-dimensional 10th nearest neighbor density, noting that the former tends to under-estimate density in high-mass galaxy clusters. It is also common practice to use the three-dimensional galaxy correlation function as a metric for density \citep[eg:][]{Zehavi_2005}, although it should be noted that the two-point correlation function is insensitive to higher order correlations which almost certainly exist between galaxies. 

We compute densities using three-dimensional positions. Rather than use the traditional tenth nearest neighbor metric for number density, $N_{10} = 1/d_{10}^{3}$, we use a Bayesian metric \citep{Ivezic_2005}: 
\begin{equation}
n = C \frac{1}{\sum_{i=1}^{10} d_{i}^{3}},  
\end{equation}
where $d_{1}=0$ if computing the density at the location of a galaxy. The constant $C=11.48$ is empirically determined by demanding that $\langle n \rangle$ matches actual number density when density is estimated on a regular grid for a uniform density field. As shown in \cite{Ivezic_2005}, the use of distances to all ten neighbors, as opposed to only the tenth neighbor, results in a factor of $\sim 2$ improvement in the precision of density estimates.  

\begin{figure}[!p]
\includegraphics[width=1\textwidth]{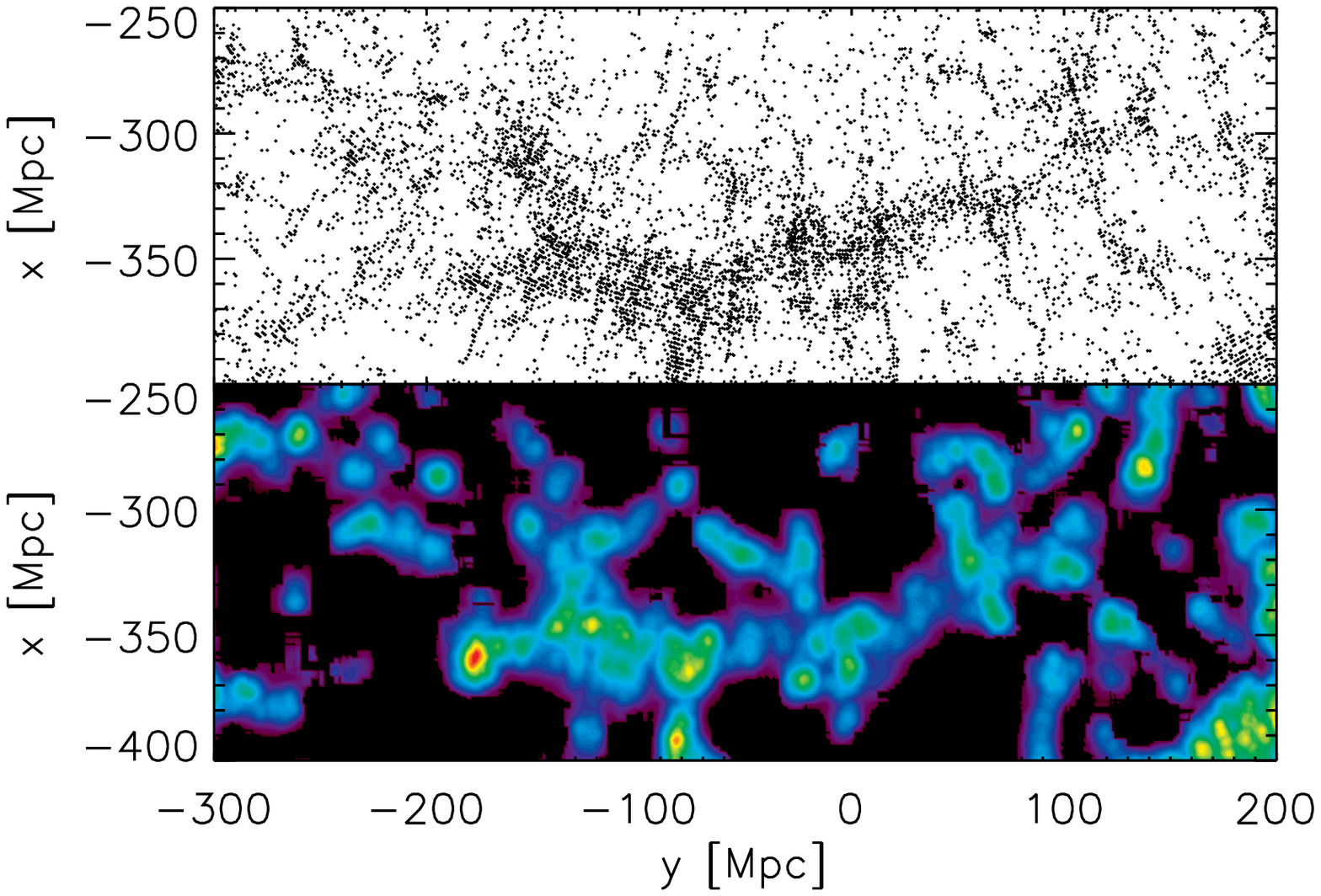}
\caption{A map of the Sloan Great Wall, located on the celestial equator some 350~Mpc away from the Milky Way \citep[see Fig.~9 in][]{Gott_2005}. Due to the large distance to the purported structure, we applied a $M_{r} <-21$ cut to the DR5 galaxy catalog, yielding a sample of 129,974 galaxies complete to $z=0.12$. The top panel shows the actual distribution of galaxies within $\pm 7^{\circ}$ of the equatorial plane. The bottom panel shows a Bayesian density map with spatial resolution of 1~Mpc with black areas corresponding to low-density environments, while red regions are the most massive clusters. This visualization resolves the wall into something more reminiscent of a mountain range.}
\label{sloan_great_wall}
\end{figure}

Although we are mostly interested in the density at the position of galaxies, the Bayesian density estimator can be used at arbitrary positions, allowing us to construct a density map on a regular grid. In Fig.~\ref{sloan_great_wall} we show the density map (with resolution 1 Mpc) for an equatorial slice of DR5, centered on the Sloan Great Wall of \cite{Gott_2005}. Thanks to the precision of our density estimator, it is easily discernible that the ``wall'' is not a monolithic structure, but results from the juxtaposition of a collection of large clusters of galaxies\footnote{With the visualization shown in Fig.~\ref{sloan_great_wall} this structure is more reminiscent of a mountain range than a great wall.}.

\section{Observational Effects} \label{biases}
Three observational effects which plague the SDSS galaxy sample could significantly affect the density distribution of galaxies: incompleteness, edge effects, and ``Fingers of God''. In this section we describe these effects and quantify how they influence the density distribution of SDSS galaxies.

For galaxies falling within 55'' of each other, only one galaxy gets a fiber, due to fiber collisions. Since 30\% of the SDSS survey area consists of overlap regions between neighboring fields, the net effect of fiber collisions is a loss of 6\% of the photometric galaxies that would otherwise be in the spectrocopic catalogue \citep{Strauss_2002}. Fiber collisions notwithstanding, more than 95\% of galaxies in the SDSS photometric catalogue are given a fiber and are in the spectroscopic catalogue. The bulk of the remaining 5\% suffer from blending with saturated stars and do not significantly bias the spectroscopic galaxy sample \citep{Strauss_2002}.

The tiling of SDSS fields is such that there are no gaps except near the edges of the survey area \citep{Blanton_2003a}. The density estimated near the edge of the DR5 footprint will be artificially low because SDSS spectra have not been obtained for many of the true nearest neighbors. To remove the most egregious offenders, we do not compute densities for any galaxies with fewer than 10 neighbors within a 10~Mpc radius or for galaxies falling within 10~Mpc of our redshift limits.

Galaxies in the SDSS spectroscopic catalog have small redshift uncertainties (30 km/sec), but the true limiting factor for determining the radial distance to galaxies is the ``Fingers of God'' effect: massive galaxy clusters have large velocity dispersion, $\sigma$, which has the effect of smearing them out in redshift-space and reducing their apparent density by a factor $1/(1+\sigma/cz)$.

To quantify the impact of observational effects, we compare the density distribution of our our mock survey of MS galaxies with that for the same survey but in which we model these effects. Fiber collisions are conservatively implemented by ignoring all neighbors within 55'' of a galaxy when computing density (this affects 6\% of the galaxies, in good agreement with the estimate for SDSS fiber collisions). General spectroscopic incompleteness is implemented by removing 5\% of the galaxies at random from the catalogue. To increase the surface area to volume ratio ---and hence the importance of edge effects--- we limit our samples to galaxies lying \textit{less than} 10~Mpc from a survey edge (reducing the size of the sample by a factor $\sim5$). We model Fingers of God by adding the peculiar velocity of each galaxy (obtained from the MS database) to its model Hubble flow, then computing its apparent radial distance from this mock redshift rather than from its actual Cartesian position. 

\begin{figure}[!p]
\includegraphics[width=1\textwidth]{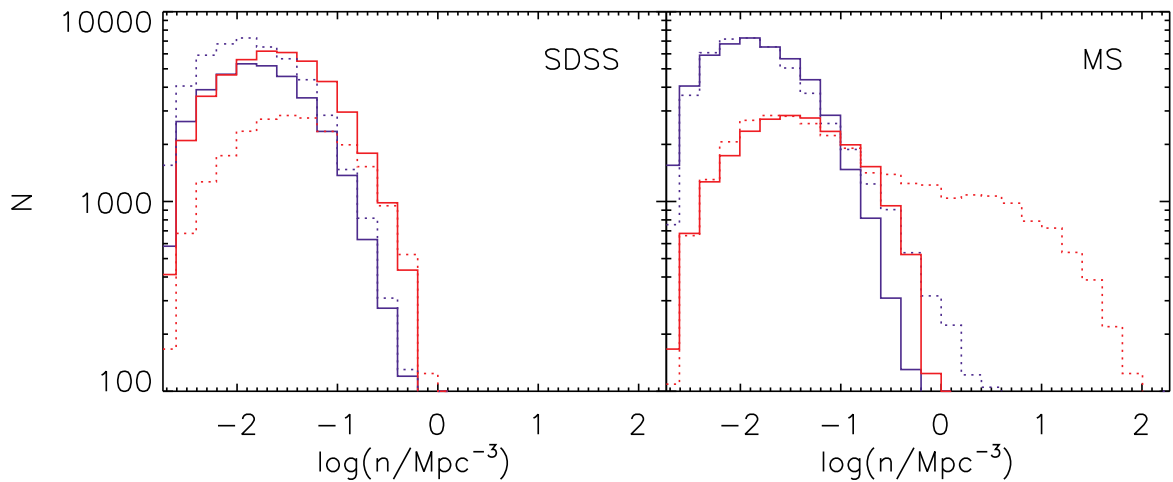}
\caption{Local number density distribution for SDSS and MS galaxies in the left and right panels, respectively. The blue and red lines represent the density distribution for blue and red galaxies based on a $u-r=2.2$ cut. For reference, the dotted lines in the left panel show the distributions for blue and red MS galaxies (same as the solid lines in the right panel). The dotted lines in the right panel represents the density distribution of the MS before the application of observational effects.}
\label{uber_density_function}
\end{figure}

The density distributions for the mock MS survey with and without observational effects are shown as solid and dotted lines in the right panel of Fig.~\ref{uber_density_function}. The distribution remains unaffected except at high densities, where Fingers of God completely erase the high density tail ($\sim 3$\% of galaxies). If this extremely high density population of galaxies exists, it will only be detected in the next generation surveys operating at higher redshifts, where the Hubble flow dominates over peculiar velocities. For the remainder of the paper we compare our SDSS galaxy sample to the mock MS survey, including the effects of fiber collisions, spectroscopic incompleteness and Fingers of God. 

\section{Environment and Photometry of Galaxies}
\begin{figure}[!p]
\includegraphics[width=1\textwidth]{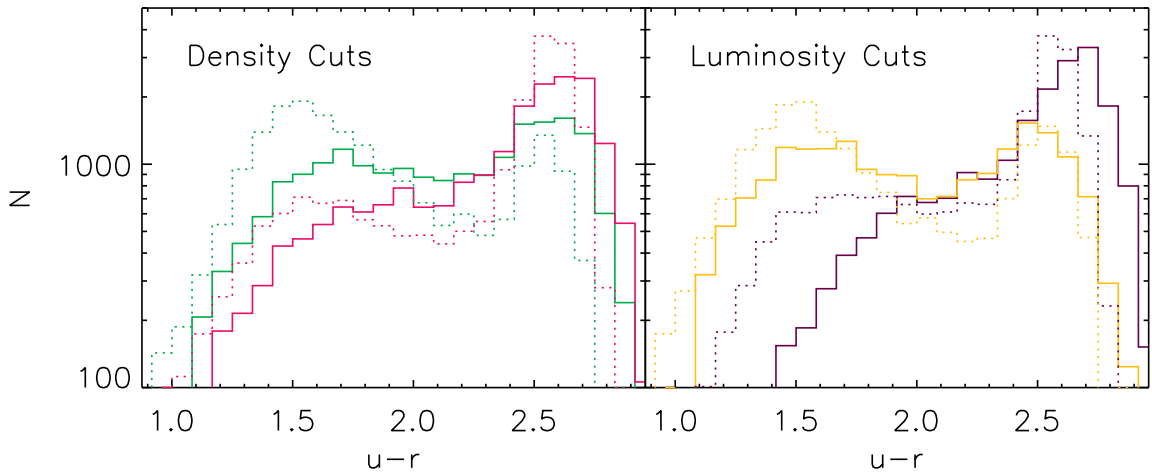}
\caption{Color distribution for SDSS (solid)  and MS (dotted) galaxies divided by density in the left panel and by luminosity in the right panel. Green and magenta lines in the left panel represent the color distribution for the lowest and highest quartiles in the density. Yellow and purple lines in the right panel represent the luminosity function for the lowest and highest quartiles in luminosity.}
\label{uber_color_function}
\end{figure}

For the purposes of plotting our results, we remove outliers from both galaxy samples by cutting out the top and bottom percentile in color, as well as the top and bottom $0.1$\% in luminosity and density. The MS histograms are rescaled to the same total number of galaxies as for the SDSS. Fig.~\ref{uber_density_function} shows the density distributions for the SDSS and MS galaxies, separated into the blue and red mode based on the $u-r = 2.2$ cut of \cite{Strateva_2001} (For comparison, the peak of the density distribution for random positions in the survey is $n = 10^{-2.8}$~Mpc$^{-3}$ in either galaxy sample). There are $50$\% too many blue galaxies in the MS as compared to the SDSS, despite the fact that the minimum in color for both samples occurs at $u-r = 2.2$ (see Fig.~\ref{uber_color_function}). 

\begin{figure}[!p]
\includegraphics[width=1\textwidth]{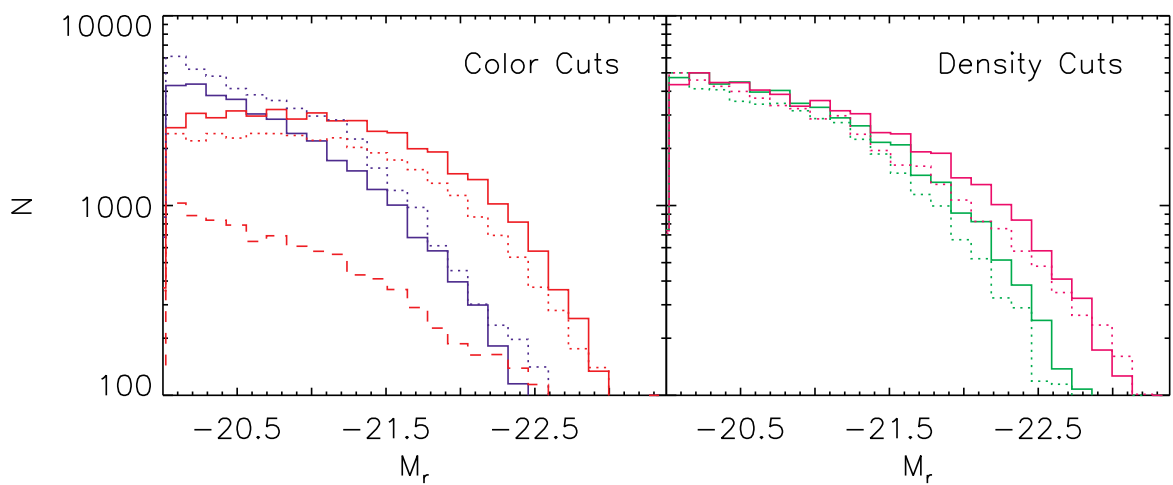}
\caption{Luminosity function for SDSS (solid) and MS (dotted) galaxies divided by color in the left panel and by density in the right panel. Red and blue lines in the left panel represent the luminosity functions for red and blue galaxies, based on a $u-r=2.2$ cut. The dashed line in the left panel shows the luminosity function of the MS galaxies with $n > 1 $~Mpc$^{-3}$ before the application of observational effects. Green and magenta lines in the right panel represent the luminosity function for the lowest and highest quartiles in density (normalized at the faint end).}
\label{uber_luminosity_function}
\end{figure}

The luminosity function for both sets of galaxies, shown in Fig.~\ref{uber_luminosity_function}, match very well except for the over-representation of blue galaxies in the MS. The luminosity function of the very dense ($n > 1$~Mpc$^-3$) modeled galaxies ---invisible in the mock survey--- is shown with the dotted line in the left panel of Fig.~\ref{uber_luminosity_function}. These extremely high density environments are populated by relatively faint ($M_{r} > -21.5$) red galaxies, not LRGs. The luminosity function for the lowest and highest density quartiles, shown in the right panel of Fig.~\ref{uber_luminosity_function}, indicates that the SAM reproduces the environmental dependence of luminosity.

Fig.~\ref{uber_color_function} shows the color distribution for the lowest and highest density quartiles of each sample. The SDSS and MS galaxy samples both exhibit a bimodal color distribution with a minimum at $u-r=2.2$, although the blue peak is too pronounced for MS galaxies. The peaks of the blue and red populations for the MS galaxies are approximately $0.2$ magnitudes too blue, as compared to the SDSS galaxies. In both panels, the relative heights of the red and blue peaks change as a function of density. Red galaxies represent $\sim 2/3$ of the highest density quartiles of both the SDSS and MS samples. The environmental dependence of color is exaggerated for the lowest-density quartile in the MS, however: 79\% are blue, compared to only 52\% for the SDSS sample \citep[see also][]{Patiri_2006}. The right panel of Fig.~\ref{uber_color_function} indicates that the  SAM fails to reproduce the dependance of color on luminosity, namely that brighter red galaxies are redder.

\begin{figure}[!p]
\includegraphics[width=1\textwidth]{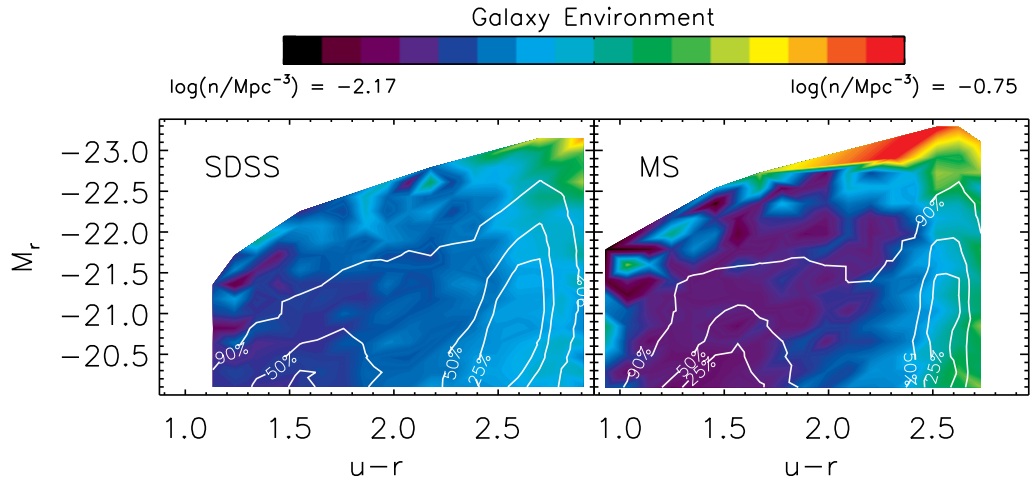}
\caption{Color magnitude diagram for SDSS and MS galaxies in the left and right panels, respectively. Labeled white lines show which regions on the plot are most populated (these are complete volume-limted samples), while the color-coded background shows the median local environment around galaxies with a given color and magnitude (dark corresponds to low densities; bright corresponds to high densities).}
\label{uber_color_magnitude}
\end{figure}

All of the features and discrepancies described above can be qualitatively seen in Fig.~\ref{uber_color_magnitude}, which combines the color, luminosity and density information for all the galaxies. Labeled white lines show which regions on the plot are most populated, while the color contours denote the median density of galaxies in a given color-magnitude bin. For galaxies less luminous than $M_{r}=-22$, $u-r$ color tracks density, while luminosity is independent of environment. For the brightest galaxies, however, density correlates with luminosity and not with color. 

\section{Conclusions}
We have compared two analogous galaxy samples, one from the SDSS DR5 spectroscopic sample, and one from the SAMs of \cite{DeLucia_2007}, after correcting for the observational effects present in the former. The density distribution and the luminosity function of the modeled galaxies qualitatively match those for the SDSS sample, but there are $50$\% too many blue galaxies in the former. In detail, two additional discrepancies become apparent between the galaxy samples: MS galaxies are more blue in $u-r$ than SDSS galaxies; the colors of galaxies depend more strongly on environment in MS than in SDSS. The strong environmental dependence manifests itself as an over-representation of blue galaxies overall and suggests that the feedbacks implemented by \cite{DeLucia_2007} exaggerate the role of galaxy environment. A population of relatively faint red galaxies in extremely high density environments is visible in the MS survey \emph{sans} observational effects. Such high density environments would be imperceptible in SDSS due to velocity dispersions comparable to local Hubble flow. 

\acknowledgments
N.B.C. is supported by the Natural Sciences and Engineering Research Council of Canada. The authors wish to thank G. Lemson for his help with MS queries, as well as R. Ro\v skar, D. Patton and the anonymous referee for invaluable comments about this work. M. Juri\' c came up with the ``mountain range'' analogy. The Millennium Simulation was carried out by the Virgo Consortium at the Computing Center of the Max Planck Society in Garching. Funding for the Sloan Digital Sky Survey (SDSS) and SDSS-II has been provided by the Alfred P. Sloan Foundation, the Participating Institutions, the National Science Foundation, the U.S. Department of Energy, the National Aeronautics and Space Administration, the Japanese Monbukagakusho, and the Max Planck Society, and the Higher Education Funding Council for England. The SDSS Web site is http://www.sdss.org/.

\clearpage

\end{document}